\documentclass[aps,twocolumn,showpacs,pra,superscriptaddress,amsmath,amssymb]{revtex4-1}
\usepackage{amsmath,amssymb,bm}
\usepackage{color}
\usepackage{graphicx}
\usepackage{dcolumn}
\usepackage{bm}
\usepackage{cases}
\usepackage[dvipdfmx]{hyperref}

\usepackage{ulem}

\begin{document}

\title{Valence Fluctuations in the Extended Anderson Lattice Model
with Quasiperiodicity}
\author{Ryu Shinzaki}
\affiliation{Department of Physics, Tokyo Institute of Technology,
Meguro-ku, Tokyo 152-8551, Japan}
\author{Joji Nasu}
\affiliation{Department of Physics, Tokyo Institute of Technology,
Meguro-ku, Tokyo 152-8551, Japan}
\author{Akihisa Koga}
\affiliation{Department of Physics, Tokyo Institute of Technology,
Meguro-ku, Tokyo 152-8551, Japan}

\date{\today}

\begin{abstract}
We study valence fluctuations in the extended Anderson model
on two-dimensional Penrose lattice, using the real-space dynamical
mean-field theory combined with the continuous-time Monte Carlo method.
Calculating $f$-electron number, $c$-$f$ spin correlations, and
magnetic susceptibility at each site,
we find site-dependent formations of the singlet state and
valence distribution at low temperatures,
which are reflected by the quasiperiodic lattice structure.
The bulk magnetic susceptibility is also addressed.
\end{abstract}

\pacs{71.10.Fd,71.23.Ft,75.20.Hr}


\maketitle
\section{Introduction}
Strongly correlated electron systems have attracted great attention.
Typical examples are rare-earth compounds, where a variety of phenomena
have been observed such as
heavy fermion behavior~\cite{PhysRevLett.35.1779,Heavy,Sato1989},
unconventional superconductivity~\cite{Steglich},
and quantum critical behavior~\cite{qc1,qc2,qc3,PhysicaB+C.91.231}.
These interesting low-temperature properties are understood by 
 the itinerancy of $f$-electrons through
the hybridization to conduction bands and
strong electron correlations, which are described by the Anderson model. 
Valence-related phenomena
have also been observed in some Ce-based and Yb-based compounds
such as valence transitions~\cite{vCe,PhysRevLett.42.321,PhysRevB.19.4154,vYb} and superconductivity~\cite{PhysRevB.30.1182,Physica.B.259.676,Jaccard19991,Science.302.5653.2104,PhysRevB.69.024508}. It has been claimed that the Coulomb repulsion between the conduction
and localized $f$-electrons play an essential role
in understanding these phenomena~\cite{Onishi}.
This stimulates further investigations on strongly correlated electron systems
with valence fluctuations~\cite{JPSJ.75.043710,JPSJ.77.024716,JPSJ.80.114711,Kojima,PhysRevB.87.125146,JPCS2016012041}.

Recently, interesting phenomena have been observed
in the $\rm Au-Al-Yb$ alloy,
which is composed of the Tsai-type clusters with an intermediate valence
in the ytterbium ions~\cite{Nmat.12.1013,JPSJ.83.034705}.
The quasicrystal, where the clusters are arranged
with quasiperiodic structure,
exhibits the divergence in the magnetic susceptibility and the specific heat with unconventional critical exponents~\cite{Nmat.12.1013}.
On the other hand, in the approximant for the $\mathrm{Au-Al-Yb}$ alloy with periodic structure,
quantum critical behavior does not appear,
but heavy-fermion behavior has been observed~\cite{Nmat.12.1013}.
These results suggest the existence of quantum critical behavior specific to the quasicrystal.
Some theoretical studies have been done to understand
the origin of the quantum critical behavior
by focusing on the structure of Tsai-type cluster~\cite{JPSJ.82.083704}
and the Kondo-disorder~\cite{PhysRevLett.115.036403}.
However, the previous experimental studies suggest that strong correlations
in $f$-electron system 
accompanied by the quasiperiodic structure are crucial
to understand the interesting low-temperature properties.
This opens a new avenue for valence fluctuations
in the heavy fermion systems.

In our previous work,
the effect of the quasiperiodic structure in
strongly correlated electron systems has been studied, by
considering a two-dimensional Penrose lattice~\cite{Takemura,Takemori}.
It has been clarified that nontrivial valence distribution emerges
at low temperatures,
by treating correlation effects by means of the non-crosssing
approximation~\cite{NCA1,NCA2}.
However, the reliable results are restricted at rather high temperatures.
Therefore, it is necessary to 
study interesting valence distributions indeed appearing at low temperatures.
Furthermore, it is instructive
to clarify how the quasiperiodic structure affects magnetic properties
in the bulk,
which may be important to discuss quantum critical behavior in the compound
$\mathrm{Au}_{51}\mathrm{Al}_{34}\mathrm{Yb}_{15}$.

In this paper, we study valence and magnetic fluctuations
in an extended Anderson lattice model.
In order to discuss the quasiperiodic structure and
local electron correlations on an equal footing,
we apply the real-space version of dynamical mean-field theory
(R-DMFT)~\cite{RevModPhys.68.13,AdvPhys.44.187,ZPhysB.74.507,PhysRevLett.62.324}
to the model on the two-dimensional Penrose lattice.
Calculating $f$-electron number,
$c$-$f$ spin correlations, and local magnetic susceptibility at each site,
we discuss low temperature properties characteristic of electron systems on
the quasiperiodic lattice.
In addition, the temperature dependence of the bulk susceptibility
is also discussed.

This paper is organized as follows.
In Sec.~\ref{MM}, we introduce the model Hamiltonian and
summarize the method used in the present study.
In Sec.~\ref{results}, we calculate valence distributions,
$c$-$f$ spin correlations and local magnetic susceptibility,
and discuss the effects of the quasiperiodic structure.
A brief summary is provided in Sec.~\ref{conclusion}.



\section{Model and Method}\label{MM}
We study valence fluctuations in the extended Anderson lattice
model~\cite{Onishi}, which is described by the following Hamiltonian:
\begin{eqnarray}
\mathcal{H}&=-t\displaystyle\sum_{\langle i,j\rangle,\sigma}
c_{i\sigma}^{\dagger}c_{j\sigma}+V\sum_{i,\sigma}
(c_{i\sigma}^{\dagger}f_{i\sigma}+\mathrm{h.c.})
+\epsilon_f\sum_{i,\sigma}n^f_{i\sigma}  \nonumber \\
&+ U_{ff}\displaystyle\sum_{i}n^f_{i\uparrow}n^f_{i\downarrow}
+ U_{cf}\sum_{i,\sigma,\sigma'}n^c_{i\sigma}n^f_{i\sigma'},
\end{eqnarray}
where $c_{i\sigma}$ ($f_{i\sigma}$) is an annihilation operator
of a conduction electron ($f$-electron)
with spin $\sigma(=\uparrow,\downarrow)$.
$n^c_{i\sigma}(=c^\dagger_{i\sigma}c_{i\sigma})$ and
$n^f_{i\sigma}(=f^\dagger_{i\sigma}f_{i\sigma})$ are
the number operators of the conduction
and $f$-electrons at $i$th site, respectively.
$t$ is the hopping integral of the conduction electrons
between the nearest-neighbor sites,
$V$ is the hybridization between the conduction band and $f$-orbitals,
and $\epsilon_f$ is the energy level of the $f$-orbitals.
$U_{ff}$ 
and $U_{cf}$ are the repulsive interactions between the $f$-electrons, 
and between the conduction and $f$-electrons, respectively.

When the system is periodic,
the Anderson lattice model has been investigated
in detail~\cite{Grewe,ohashi_field-induced,PhysRevLett.99.196403,koga_2010}.
In the case of $\epsilon_f \ll -U_{ff}/2$,
the strong Coulomb interaction $U_{ff}$
tends to fix the $f$-electron number as unity.
At higher temperatures, the local moment state with a free $f$-electron spin is realized at each site.
At low temperatures, local spins are screened by conduction electrons
and the Kondo singlet state is realized.
In the large $\epsilon_f$ case, the $f$-elctron number
is away from unity and the mixed-valence state is realized.
When $U_{cf}=0$, the mixed-valence and Kondo singlet states
are adiabatically connected to each other and
the valence crossover occurs when $\epsilon_f$ is changed.
By contrast, the introduction of $U_{cf}$ enhances
valence fluctuations,
which leads to the first-order valence transition~\cite{Onishi,JPSJ.75.043710,JPSJ.77.024716,JPSJ.80.114711,Kojima,PhysRevB.87.125146,JPCS2016012041}.

\begin{figure}[tb]
  \centering
  \includegraphics[width=5.5cm]{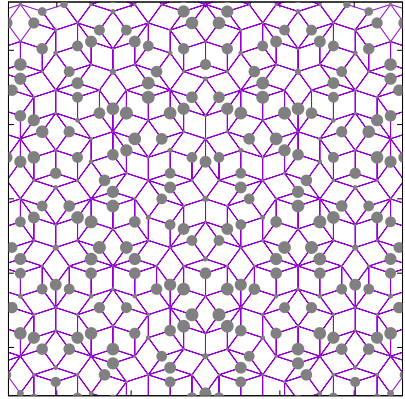}
\caption{
 Two dimensional Penrose lattice.
The radii of circles represent the $c-f$ spin correlations
$|\langle{\bf S}_c\cdot {\bf S}_f\rangle|$ in a certain parameter space (see text).
}
\label{Lattice}
\end{figure}

In the paper, we discuss how valence fluctuations
affect valence distribution and spin correlations in a quasiperiodic system.
To this end,
we consider the two-dimensional Penrose lattice,
which consists of fat and skinny rhombuses (see Fig.~\ref{Lattice}).
One of the important features is that the Penrose lattice has
the five-fold rotational symmetry.
Therefore, in our paper, we treat the Penrose lattice
with open boundary and five-fold rotational symmetry,
which is iteratively generated in terms of the inflation-deflation rule~\cite{PhysRevLett.53.2477}.

We study the extended Anderson Hamiltonian
for the vertex model of the Penrose lattice,
where the sites are placed on the corner of rhombuses.
The coordination number $Z$ ranges from three to seven
except for the edge sites.
To treat site-dependent behavior correctly,
we use the R-DMFT~\cite{RevModPhys.68.13,AdvPhys.44.187,ZPhysB.74.507,PhysRevLett.62.324},
where local electron correlations are taken into account.
The method has successfully been applied to inhomogeneous systems
such as the surface~\cite{Potthoff}, interface~\cite{Okamoto},
superlattice~\cite{Peters},
optical lattice~\cite{Snoek,Helmes,KogaOL,KogaOL2},
topological insulator~\cite{Tada}.
In the framework of the R-DMFT, an effective impurity model is
constructed at each site.
The selfenergy $\mathbf{\Sigma}_{\sigma}$
is assumed to be site-diagonal in this formalism.
This allows us to calculate the lattice Green's function
$\mathbf{G}_\sigma$ as
\begin{eqnarray}
\mathbf{G}_\sigma^{-1}=\mathbf{G}_{0 \sigma}^{-1}-\mathbf{\Sigma}_{\sigma},
\end{eqnarray}
where $\mathbf{G}_{0 \sigma}$ is non-interacting lattice Green's function,
which is given by
\begin{eqnarray}
\mathbf{G}_{0 \sigma}^{-1}
=
\begin{pmatrix}
i\omega_n+\mu & -V \\
-V & i\omega_n+\mu-\epsilon_f \\
\end{pmatrix}
\delta_{i j} 
+
\begin{pmatrix}
t& 0 \\
0 & 0 \\
\end{pmatrix}
\delta_{\langle ij\rangle},
\end{eqnarray}
where $\omega_n\equiv(2n+1)\pi T$ is the Matsubara frequency
with temperature $T$ and integer $n$, and $\mu$ is the chemical potential.
$\delta_{\langle ij\rangle}$ is 1 when site $i$ and $j$ are neighboring sites and zero otherwise.
The Green's function of an effective bath at the $i$th site
is determined as
\begin{eqnarray}
\mathcal{G}_{\sigma}^{(i)} (i\omega_n)^{-1}=
[\mathbf{G}^{-1}_{\sigma}(i\omega_n)]_{ii}
+\mathbf{\Sigma}_{i \sigma}(i\omega_n).
\end{eqnarray}
The above self-consistent calculations are iterated
until the Green's functions of the effective bath are converged.
This method enables us to deal with local electron correlations
and the 
quasiperiodicity on an equal footing.

In order to solve the effective impurity models at each site,
we make use of the continuous-time quantum Monte Carlo method
based on the hybridization expansion~\cite{PhysRevLett.97.076405,PhysRevB.74.155107,RevModPhys.83.349}.
The partition function of the effective impurity model is expanded
in powers of the mixing between the impurity site and effective bath.
This enables us to evaluate physical quantities quantitatively
in terms of the Monte Carlo procedure,
contrast to other biased methods such as the iterative perturbation
theory~\cite{IPT} and non-crossing approximation~\cite{NCA1,NCA2}.

To discuss low temperature properties in the quasiperiodic lattice,
we calculate the local quantities such as
$f$-electron number $\langle n_i^f\rangle$ and $c-f$ spin correlations
$\langle {\bf S}_c\cdot {\bf S}_f\rangle$.
We also calculate the local susceptibility
\begin{eqnarray}
\chi_{i}&=& \int_0^{\beta}d \tau
\langle M_{z,i}(\tau)M_{z,i} \rangle_{\mathrm{imp}},
\end{eqnarray}
where $M_{z,i} = n^c_\uparrow-n^c_\downarrow+n^f_\uparrow-n^f_\downarrow$
is the local magnetic moment operator.
For simplicity, the total susceptibility is assumed to be given as
$\chi=\sum_{i}\chi_{i}/N$,
where $N$ is the total number of sites.

In the paper, we use $t$ as the unit of energy.
We choose the parameters as $U_{ff}/t=20$, $U_{cf}/t=5$, and $N=601$, and
the total electron number per site is
fixed as $\langle n^c\rangle + \langle n^f\rangle=1.9$, where $\langle n^c \rangle \equiv \sum_i\langle n_i^c\rangle/N$ and $\langle n^f \rangle \equiv \sum_i\langle n_i^f\rangle/N$.

\section{results}\label{results}

\begin{figure}[tb]
  \centering
  \includegraphics[width=7.5cm]{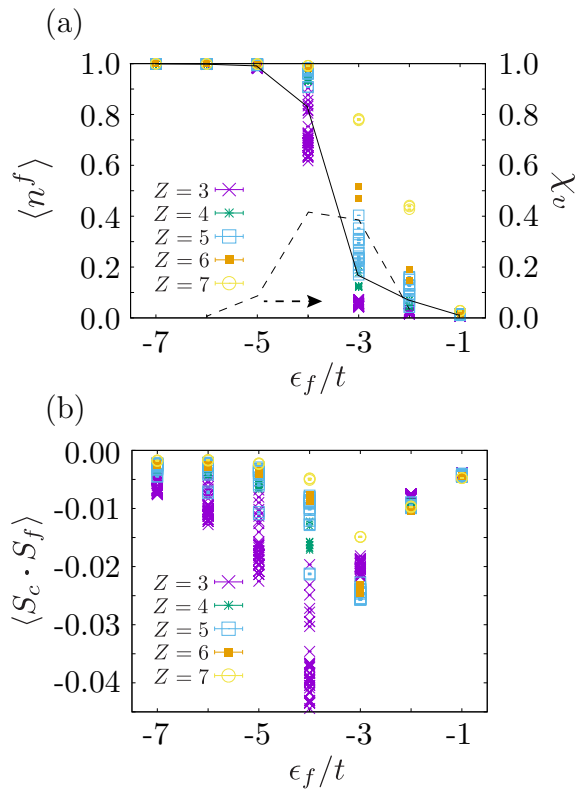}
\caption{
$f$-energy level dependence of (a) the number of $f$-electrons and
(b) the $c-f$ spin correlations at each site in the system with $V/t=0.2$ and $T/t=0.2$.
Data are classified into five groups in terms of the coordination number $Z$.
Solid (Dashed) line in (a) represents the site average of the number of $f$-electrons and valence fluctuations $\chi_v$.}
\label{NfSCef}
\end{figure}

First, we calculate local physical quantities to discuss
the effect of the quasiperiodic structure at low temperatures.
Since the Penrose lattice used in the present calculations does not have translational symmetry,
each site is not equivalent.
Therefore, local quantities, in general, depend on the lattice site.
Figure~\ref{NfSCef} shows
the number of $f$-electrons and $c-f$ spin correlations at each site
in the system with $V/t=0.2$ at $T/t=0.2$.
When the $f$-energy level is low enough $(\epsilon_f/t=-7)$,
the $f$-electron number is fixed as $n_f\sim 1$
due to the strong Coulomb interaction $U_{ff}$.
In the case, the $c-f$ spin correlations are relatively small, indicating that Kondo singlet correlations do not develop.
Thus, we can say that the local moment state is realized in the system.
Note that, in the case, the quasiperiodic structure has little effect
on low temperature properties since the weak site-dependence appears in the local $f$ electron number and $c-f$ spin correlations.
With increasing $\epsilon_f$, 
 the $f$-electron number
$\langle n^f\rangle$ rapidly decreases and valence fluctuations
$\chi_v \equiv |d\langle n^f\rangle/d\epsilon_f|$ are enhanced
around $\epsilon^f/t \sim -4$, as shown in Fig.~\ref{NfSCef}(a).
The absence of the singularity in local quantities implies that
 the valence crossover occurs from the local moment state to the mixed-valence state with
$\langle n^f_i \rangle<1$.
What is the most important is that local quantities at each site are not identical and strongly depend on the coordination number.
For example, in the case with $\epsilon_f/t=-4$,
the sites with large coordination number $Z>5$ still
have local moments $(\langle n^f_i \rangle\sim 1)$,
while the intermediate-valence state is realized in the others, as shown in Fig.~\ref{NfSCef}(a).
This yields complex behavior in the $c-f$ correlations,
as shown in Fig.~\ref{NfSCef}(b).
Although its absolute value is rather small,
site-dependent behavior clearly appears as a characteristic of
the quasiperiodic structure shown in Fig.~\ref{Lattice}.
Further increase of $\epsilon_f$ decreases the number of $f$-electrons.
Finally, the $f$-level is almost empty with $\langle n^f_i \rangle\sim 0$ and
$\langle {\bf S}_c\cdot {\bf S}_f\rangle \sim 0$, where
the feature of the quasiperiodic structure smears.
As discussed above, this structure plays a crucial role
for low temperature properties when valence fluctuations are enhanced.

In our calculations, we could not find
the first-order valence transition in the extended Anderson model
on the Penrose lattice
although it occurs in the periodic lattice~\cite{JPSJ.75.043710}.
This may be explained as follows.
In the model, the bare onsite interactions $U_{ff}$ and $U_{cf}$ are
uniform, but the local geometry depends on the site
on the Penrose lattice.
Therefore, the site-dependent potential is effectively induced,
leading to the valence crossover with a smooth change in the valence.
This is qualitatively consistent with the results obtained by means of
DMFT with non-crossing approximations~\cite{Takemura}.

Next, we discuss how site-dependent properties appear at finite temperatures.
\begin{figure*}[tb]
  \centering
  \includegraphics[width=13cm]{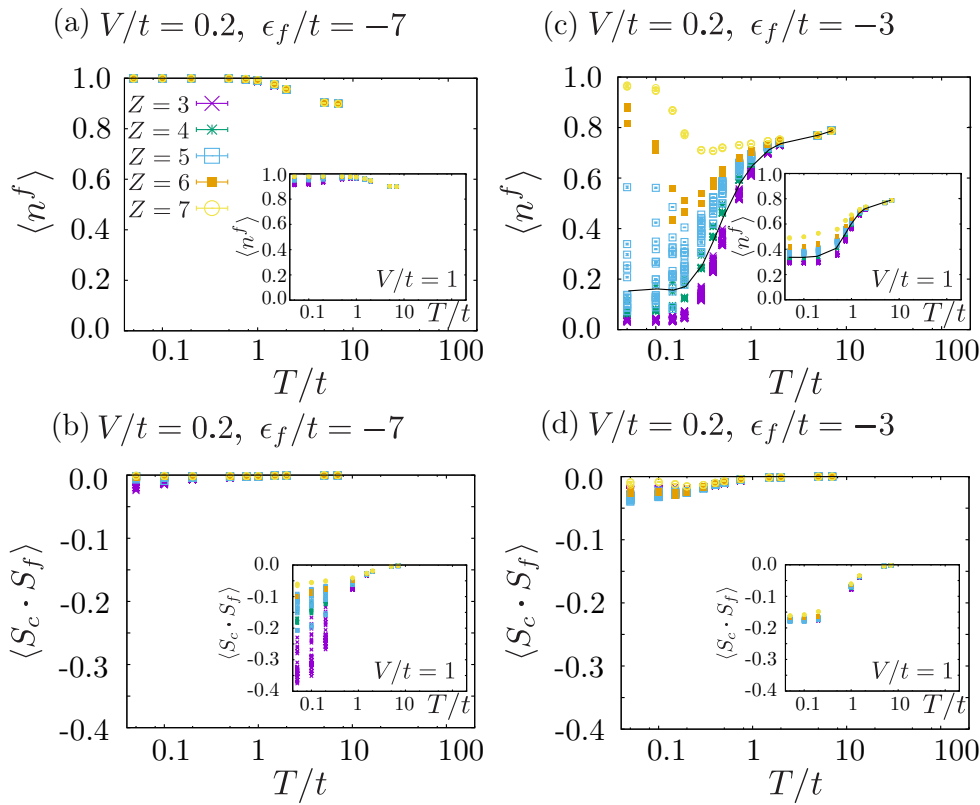}
\caption{
Temperature dependence of the number of $f$-electrons (upper panels) and $c-f$ spin correlations (lower panels) in the system with $\epsilon_{f}/t=-7$ and $\epsilon_{f}/t=-3$ at $V/t=0.2$. Symbols represent the results classified by the coordination number $Z$. Inset of the each panel shows the result at $V/t=1$.
}
\label{NfT}
\end{figure*}
Figure~\ref{NfT} shows the temperature dependence of the $f$-electron number
and $c-f$ spin correlations in the system with $V/t=0.2$.
When $\epsilon_f=-7$, $\langle n^f_i \rangle$ is almost unity and
the $c-f$ correlations are little enhanced, as shown in Figs.~\ref{NfT}(a) and~\ref{NfT}(b).
Therefore, down to $T/t=0.05$, the local moment state is realized.
If the hybridization is large, the local $f$-electron moment is screened
by the conduction electrons at low temperatures.
In fact, when $V/t=1$,
the $c-f$ correlations are enhanced at $T/t\lesssim 1$,
as shown in the inset of Fig.~\ref{NfT}(b).
It is also found that the large site-dependence appears
at low temperatures, in contrast to the distribution of
$\langle n^f_i \rangle$.
This implies that a single occupied state is realized in each $f$-orbital
and the effective Kondo temperature depends on the lattice sites.

When $\epsilon_f/t=-3$,
an interesting temperature dependence appears,
as shown in Figs.~\ref{NfT}(c) and~\ref{NfT}(d).
At high temperature $T/t=7$, $\langle n^f_i \rangle\sim 0.8$ and
the valence is distributed uniformly in the system.
Therefore, in the case, the quasiperiodic structure does not contribute to the valence distribution.
However, the decrease in temperature induces site-dependent behavior.
When $0.75 < T/t < 1.5$,
the quantity $\langle n^f_i \rangle$ is split into five groups,
which are classified by the coordination number $Z$.
With further decrease in temperature,
$\langle n^f_i \rangle$ becomes split into many groups, where
the local geometry beyond the nearest-neighbor sites
affects the electronic state~\cite{Takemura}.
When $T/t \lesssim 0.1$,
the $f$-electron number ranges from zero to unity,
while spin correlations little develop.
This implies
the coexistence of the local moment sites with $\langle n^f_i \rangle\sim 1$
and the mixed-valence sites with $\langle n_i^f\rangle<1$.
The above temperature dependence may be explained by the fact that
longer-range electron correlations develop via the hopping of
the conduction electrons with decreasing temperature.
By contrast, when the system has large hybridization, the correlation length
becomes shorter since the local Kondo singlet should be formed at each site.
In fact,
$\langle n^f_i \rangle$ almost depends only on the coordination number
when $V/t=1$, as shown in the inset of Fig.~\ref{NfT}(c).
Thus, the quasiperiodic structure plays a minor role in the state.

\begin{figure*}[tb]
  \centering
  \includegraphics[width=13cm]{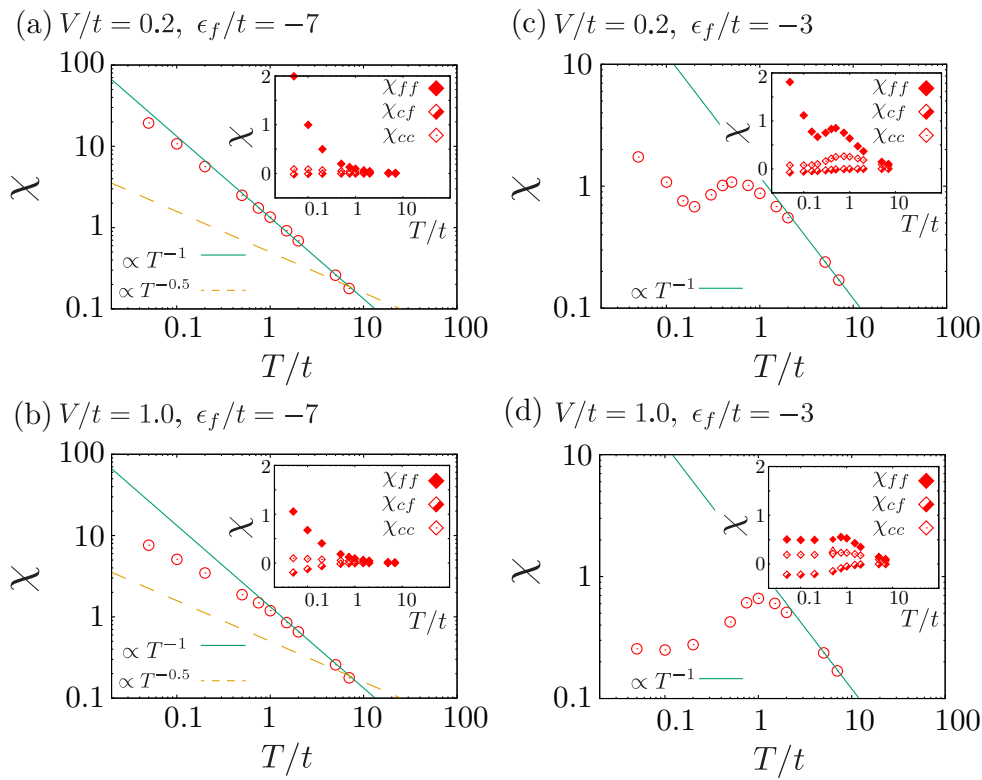}
 \caption{
Magnetic susceptibilities in the systems with at (a) $V/t=0.2, \epsilon_{f}/t=-7$, (b) $V/t=1, \epsilon_{f}/t=-7$, (c)$V/t=0.2, \epsilon_{f}/t=-3$ and (d) $V/t=1, \epsilon_{f}/t=-3$.
Inset of each panel shows the $f$-$f$, $c$-$f$, and $c$-$c$ components of the susceptibility separately.
}
\label{magsusT}
\end{figure*}

Let us consider how site-dependent behavior characteristic of
the quasiperiodic stucture affects bulk properties
in the system.
The temperature dependence of the magnetic susceptibility is shown in Fig.~\ref{magsusT}.
In the model with large $U_{ff}$, the $f$-electrons tend to be localized
in the system. Therefore, the $f-f$ component of the susceptibility
mainly contributes to the total one, while the $c-f$ and $c-c$ components
play a minor role, as shown in the insets of Fig.~\ref{magsusT}.
When $\epsilon_f/t=-7$ shown in Figs.~\ref{magsusT}(a) and (b), the local moment state is realized
at higher temperatures
and the Kondo singlet state is realized at lower temperatures,
as discussed above.
In the case $V/t=0.2$, the Kondo temperature is low enough, and
the local moment state is realized down to $T/t\sim 0.05$.
Then, the magnetic susceptibility almost obeys the Curie law
$\chi \propto 1/T$, as shown in Fig.~\ref{magsusT}(a).
When $V/t=1$, similar behavior appears at higher temperatures.
However, at low temperatures,
the Kondo singlet state is locally realized,
which suppresses the magnetic susceptibility, as shown in Fig.~\ref{magsusT}(b).

\begin{figure}[tb]
  \centering
  \includegraphics[width=7cm]{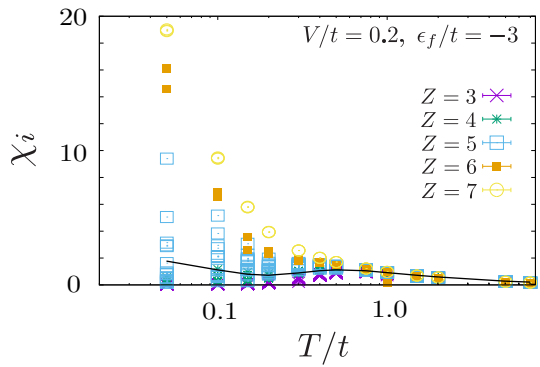}
\caption{Temperature dependence of the local magnetic susceptibility at each site when $\epsilon_{f}/t=-3$ and $V/t=0.2$. Data are classified into the coordination number $Z$ of the site.}
\label{LocalChi}
\end{figure}

On the other hand, in the mixed-valence region $(\epsilon_f/t=-3)$,
nonmonotonic behavior appears
in the magnetic susceptibility, as shown in Figs.~\ref{magsusT}(c) and~\ref{magsusT}(d).
These may be explained by the change in the $f$-electorn number.
At high temperatures,
relatively large number of electrons exist in the $f$-level.
Therefore, the susceptibility increases with decrease of temperature
in the region $T/t > 0.5$.
We find that the susceptibility has a maximum around $T/t = 0.5$
and decreases.
This should originate from the decrease of the $f$-electron number,
as shown in Fig.~\ref{NfT}(c).
Further decrease of temperature $(T/t\sim 0.2)$,
the susceptibility increases again.
In the case, its spatial average $\langle n_f\rangle$ is little changed,
while nonmonotonic behavior appears in $\langle n_i^f\rangle$ for
some sites with $Z\ge 5$.
This mainly enhances magnetic fluctuations at low temperatures.
In fact, the local susceptibilities for the corresponding sites take large values at low temperatures,
as shown in Fig.~\ref{LocalChi}.
Further decrease in temperature enhances the $c-f$ spin correlations and
the Kondo singlet state becomes stable in these sites,
where magnetic fluctuations are suppressed.
The large hybridization ($V/t=1$) tends to form the local Kondo singlet,
as discussed before.
Then, the $f$-electron valence distributes in a small range,
as shown in the inset of Fig.~\ref{NfT}(c).

Even in conventional periodic systems,
nonmonotonic behavior appears in the magnetic susceptibility,
depending on the parameters.
In the case, the magnetic susceptibilty is uniformly
enhanced at low temperatures, reflected by the average of
the $f$-electron number, in contrast to that in the quasiperiodic system.
Therefore,
it may be hard to discuss how the quasiperiodic structure affects
the bulk magnetic susceptibility.
Nevertheless, we can say that, in the Penrose lattice,
the $f$-electron number at each site depends on the local geometry
as well as the temperature,
which yields nontrivial temperature dependence
in the magnetic susceptibility.


\section{Conclusion}\label{conclusion}

We have studied the extended Anderson model
on the two dimensional Penrose lattice,
using the real-space dynamical mean-field theory with
the continuous-time quantum Monte Carlo Method.
we have clarified site-dependent formations of the singlet state and
valence distribution at low temperatures,
which is reflected by the quasiperiodic lattice structure.
We have also discussed how the local properties characteristic of
the quasiperiodic structure affects the bulk magnetic susceptibility.


\begin{acknowledgements}
We would like to thank N. Takemori for valuable discussions.
Parts of the numerical calculations were performed
in the supercomputing systems in ISSP, the University of Tokyo.
This work was partly supported by the Grant-in-Aid for Scientific
Research from JSPS, KAKENHI Grant Number 25800193,
16H01066 (A.K.), and 16K17747 (J.N.).
The  simulations have been performed using some of
the ALPS libraries~\cite{alps2}.
\end{acknowledgements}

\bibliography{refs}

\begin{thebibliography}{54}%
\makeatletter
\providecommand \@ifxundefined [1]{%
 \@ifx{#1\undefined}
}%
\providecommand \@ifnum [1]{%
 \ifnum #1\expandafter \@firstoftwo
 \else \expandafter \@secondoftwo
 \fi
}%
\providecommand \@ifx [1]{%
 \ifx #1\expandafter \@firstoftwo
 \else \expandafter \@secondoftwo
 \fi
}%
\providecommand \natexlab [1]{#1}%
\providecommand \enquote  [1]{``#1''}%
\providecommand \bibnamefont  [1]{#1}%
\providecommand \bibfnamefont [1]{#1}%
\providecommand \citenamefont [1]{#1}%
\providecommand \href@noop [0]{\@secondoftwo}%
\providecommand \href [0]{\begingroup \@sanitize@url \@href}%
\providecommand \@href[1]{\@@startlink{#1}\@@href}%
\providecommand \@@href[1]{\endgroup#1\@@endlink}%
\providecommand \@sanitize@url [0]{\catcode `\\12\catcode `\$12\catcode
  `\&12\catcode `\#12\catcode `\^12\catcode `\_12\catcode `\%12\relax}%
\providecommand \@@startlink[1]{}%
\providecommand \@@endlink[0]{}%
\providecommand \url  [0]{\begingroup\@sanitize@url \@url }%
\providecommand \@url [1]{\endgroup\@href {#1}{\urlprefix }}%
\providecommand \urlprefix  [0]{URL }%
\providecommand \Eprint [0]{\href }%
\providecommand \doibase [0]{http://dx.doi.org/}%
\providecommand \selectlanguage [0]{\@gobble}%
\providecommand \bibinfo  [0]{\@secondoftwo}%
\providecommand \bibfield  [0]{\@secondoftwo}%
\providecommand \translation [1]{[#1]}%
\providecommand \BibitemOpen [0]{}%
\providecommand \bibitemStop [0]{}%
\providecommand \bibitemNoStop [0]{.\EOS\space}%
\providecommand \EOS [0]{\spacefactor3000\relax}%
\providecommand \BibitemShut  [1]{\csname bibitem#1\endcsname}%
\let\auto@bib@innerbib\@empty
\bibitem [{\citenamefont {Andres}\ \emph {et~al.}(1975)\citenamefont {Andres},
  \citenamefont {Graebner},\ and\ \citenamefont {Ott}}]{PhysRevLett.35.1779}%
  \BibitemOpen
  \bibfield  {author} {\bibinfo {author} {\bibfnamefont {K.}~\bibnamefont
  {Andres}}, \bibinfo {author} {\bibfnamefont {J.~E.}\ \bibnamefont
  {Graebner}}, \ and\ \bibinfo {author} {\bibfnamefont {H.~R.}\ \bibnamefont
  {Ott}},\ }\href {\doibase 10.1103/PhysRevLett.35.1779} {\bibfield  {journal}
  {\bibinfo  {journal} {Phys. Rev. Lett.}\ }\textbf {\bibinfo {volume} {35}},\
  \bibinfo {pages} {1779} (\bibinfo {year} {1975})}\BibitemShut {NoStop}%
\bibitem [{\citenamefont {Stewart}(1984)}]{Heavy}%
  \BibitemOpen
  \bibfield  {author} {\bibinfo {author} {\bibfnamefont {G.~R.}\ \bibnamefont
  {Stewart}},\ }\href@noop {} {\bibfield  {journal} {\bibinfo  {journal} {Rev.
  Mod. Phys.}\ }\textbf {\bibinfo {volume} {56}},\ \bibinfo {pages} {755}
  (\bibinfo {year} {1984})}\BibitemShut {NoStop}%
\bibitem [{\citenamefont {Satoh}\ \emph {et~al.}(1989)\citenamefont {Satoh},
  \citenamefont {Fujita}, \citenamefont {Maeno}, \citenamefont {Onuki},\ and\
  \citenamefont {Komatsubara}}]{Sato1989}%
  \BibitemOpen
  \bibfield  {author} {\bibinfo {author} {\bibfnamefont {K.}~\bibnamefont
  {Satoh}}, \bibinfo {author} {\bibfnamefont {T.}~\bibnamefont {Fujita}},
  \bibinfo {author} {\bibfnamefont {Y.}~\bibnamefont {Maeno}}, \bibinfo
  {author} {\bibfnamefont {Y.}~\bibnamefont {Onuki}}, \ and\ \bibinfo {author}
  {\bibfnamefont {T.}~\bibnamefont {Komatsubara}},\ }\href@noop {} {\bibfield
  {journal} {\bibinfo  {journal} {J. Phys. Soc. Jpn.}\ }\textbf {\bibinfo
  {volume} {58}},\ \bibinfo {pages} {1012} (\bibinfo {year}
  {1989})}\BibitemShut {NoStop}%
\bibitem [{\citenamefont {Steglich}\ \emph {et~al.}(1979)\citenamefont
  {Steglich}, \citenamefont {Aarts}, \citenamefont {Bredl}, \citenamefont
  {Lieke}, \citenamefont {Meschede}, \citenamefont {Franz},\ and\ \citenamefont
  {Sch\"afer}}]{Steglich}%
  \BibitemOpen
  \bibfield  {author} {\bibinfo {author} {\bibfnamefont {F.}~\bibnamefont
  {Steglich}}, \bibinfo {author} {\bibfnamefont {J.}~\bibnamefont {Aarts}},
  \bibinfo {author} {\bibfnamefont {C.~D.}\ \bibnamefont {Bredl}}, \bibinfo
  {author} {\bibfnamefont {W.}~\bibnamefont {Lieke}}, \bibinfo {author}
  {\bibfnamefont {D.}~\bibnamefont {Meschede}}, \bibinfo {author}
  {\bibfnamefont {W.}~\bibnamefont {Franz}}, \ and\ \bibinfo {author}
  {\bibfnamefont {H.}~\bibnamefont {Sch\"afer}},\ }\href {\doibase
  10.1103/PhysRevLett.43.1892} {\bibfield  {journal} {\bibinfo  {journal}
  {Phys. Rev. Lett.}\ }\textbf {\bibinfo {volume} {43}},\ \bibinfo {pages}
  {1892} (\bibinfo {year} {1979})}\BibitemShut {NoStop}%
\bibitem [{\citenamefont {Lohneysen}(1996)}]{qc1}%
  \BibitemOpen
  \bibfield  {author} {\bibinfo {author} {\bibfnamefont {A.}~\bibnamefont
  {Lohneysen}},\ }\href@noop {} {\bibfield  {journal} {\bibinfo  {journal} {J.
  Phys.: Condens. Matter}\ }\textbf {\bibinfo {volume} {8}},\ \bibinfo {pages}
  {9689} (\bibinfo {year} {1996})}\BibitemShut {NoStop}%
\bibitem [{\citenamefont {Stewart}(2001)}]{qc2}%
  \BibitemOpen
  \bibfield  {author} {\bibinfo {author} {\bibfnamefont {G.~R.}\ \bibnamefont
  {Stewart}},\ }\href@noop {} {\bibfield  {journal} {\bibinfo  {journal} {Rev.
  Mod. Phys.}\ }\textbf {\bibinfo {volume} {73}},\ \bibinfo {pages} {797}
  (\bibinfo {year} {2001})}\BibitemShut {NoStop}%
\bibitem [{\citenamefont {Si}\ \emph {et~al.}(2000)\citenamefont {Si},
  \citenamefont {Rabello}, \citenamefont {Ingersent},\ and\ \citenamefont
  {Smith}}]{qc3}%
  \BibitemOpen
  \bibfield  {author} {\bibinfo {author} {\bibfnamefont {Q.}~\bibnamefont
  {Si}}, \bibinfo {author} {\bibfnamefont {S.}~\bibnamefont {Rabello}},
  \bibinfo {author} {\bibfnamefont {K.}~\bibnamefont {Ingersent}}, \ and\
  \bibinfo {author} {\bibfnamefont {J.~L.}\ \bibnamefont {Smith}},\ }\href@noop
  {} {\bibfield  {journal} {\bibinfo  {journal} {Nature}\ }\textbf {\bibinfo
  {volume} {413}},\ \bibinfo {pages} {804} (\bibinfo {year}
  {2000})}\BibitemShut {NoStop}%
\bibitem [{\citenamefont {Doniach}(1977)}]{PhysicaB+C.91.231}%
  \BibitemOpen
  \bibfield  {author} {\bibinfo {author} {\bibfnamefont {S.}~\bibnamefont
  {Doniach}},\ }\href@noop {} {\bibfield  {journal} {\bibinfo  {journal}
  {Physica B+C}\ }\textbf {\bibinfo {volume} {91}},\ \bibinfo {pages} {231}
  (\bibinfo {year} {1977})}\BibitemShut {NoStop}%
\bibitem [{\citenamefont {Gschneidner}\ and\ \citenamefont
  {Eyring}(1978)}]{vCe}%
  \BibitemOpen
  \bibfield  {author} {\bibinfo {author} {\bibfnamefont {K.~A.}\ \bibnamefont
  {Gschneidner}}\ and\ \bibinfo {author} {\bibfnamefont {L.}~\bibnamefont
  {Eyring}},\ }\href@noop {} {\bibfield  {journal} {\bibinfo  {journal}
  {{\it{Handbook on the Physics and Chemistry of Rare Earths}}}\ } (\bibinfo
  {year} {1978})}\BibitemShut {NoStop}%
\bibitem [{\citenamefont {Manheimer}\ and\ \citenamefont
  {Parks}(1979)}]{PhysRevLett.42.321}%
  \BibitemOpen
  \bibfield  {author} {\bibinfo {author} {\bibfnamefont {M.~A.}\ \bibnamefont
  {Manheimer}}\ and\ \bibinfo {author} {\bibfnamefont {R.~D.}\ \bibnamefont
  {Parks}},\ }\href {\doibase 10.1103/PhysRevLett.42.321} {\bibfield  {journal}
  {\bibinfo  {journal} {Phys. Rev. Lett.}\ }\textbf {\bibinfo {volume} {42}},\
  \bibinfo {pages} {321} (\bibinfo {year} {1979})}\BibitemShut {NoStop}%
\bibitem [{\citenamefont {Jayaraman}\ and\ \citenamefont
  {Maines}(1979)}]{PhysRevB.19.4154}%
  \BibitemOpen
  \bibfield  {author} {\bibinfo {author} {\bibfnamefont {A.}~\bibnamefont
  {Jayaraman}}\ and\ \bibinfo {author} {\bibfnamefont {R.~G.}\ \bibnamefont
  {Maines}},\ }\href {\doibase 10.1103/PhysRevB.19.4154} {\bibfield  {journal}
  {\bibinfo  {journal} {Phys. Rev. B}\ }\textbf {\bibinfo {volume} {19}},\
  \bibinfo {pages} {4154} (\bibinfo {year} {1979})}\BibitemShut {NoStop}%
\bibitem [{\citenamefont {Felner}\ and\ \citenamefont {Nowik}(1986)}]{vYb}%
  \BibitemOpen
  \bibfield  {author} {\bibinfo {author} {\bibfnamefont {I.}~\bibnamefont
  {Felner}}\ and\ \bibinfo {author} {\bibfnamefont {I.}~\bibnamefont {Nowik}},\
  }\href {\doibase 10.1103/PhysRevB.33.617} {\bibfield  {journal} {\bibinfo
  {journal} {Phys. Rev. B}\ }\textbf {\bibinfo {volume} {33}},\ \bibinfo
  {pages} {617} (\bibinfo {year} {1986})}\BibitemShut {NoStop}%
\bibitem [{\citenamefont {Bellarbi}\ \emph {et~al.}(1984)\citenamefont
  {Bellarbi}, \citenamefont {Benoit}, \citenamefont {Jaccard}, \citenamefont
  {Mignot},\ and\ \citenamefont {Braun}}]{PhysRevB.30.1182}%
  \BibitemOpen
  \bibfield  {author} {\bibinfo {author} {\bibfnamefont {B.}~\bibnamefont
  {Bellarbi}}, \bibinfo {author} {\bibfnamefont {A.}~\bibnamefont {Benoit}},
  \bibinfo {author} {\bibfnamefont {D.}~\bibnamefont {Jaccard}}, \bibinfo
  {author} {\bibfnamefont {J.~M.}\ \bibnamefont {Mignot}}, \ and\ \bibinfo
  {author} {\bibfnamefont {H.~F.}\ \bibnamefont {Braun}},\ }\href {\doibase
  10.1103/PhysRevB.30.1182} {\bibfield  {journal} {\bibinfo  {journal} {Phys.
  Rev. B}\ }\textbf {\bibinfo {volume} {30}},\ \bibinfo {pages} {1182}
  (\bibinfo {year} {1984})}\BibitemShut {NoStop}%
\bibitem [{\citenamefont {Miyake}\ \emph {et~al.}(1999)\citenamefont {Miyake},
  \citenamefont {Narikiyo},\ and\ \citenamefont {Onishi}}]{Physica.B.259.676}%
  \BibitemOpen
  \bibfield  {author} {\bibinfo {author} {\bibfnamefont {K.}~\bibnamefont
  {Miyake}}, \bibinfo {author} {\bibfnamefont {O.}~\bibnamefont {Narikiyo}}, \
  and\ \bibinfo {author} {\bibfnamefont {Y.}~\bibnamefont {Onishi}},\ }\href
  {\doibase http://dx.doi.org/10.1016/S0921-4526(98)00754-6} {\bibfield
  {journal} {\bibinfo  {journal} {Physica B: Condensed Matter}\ }\textbf
  {\bibinfo {volume} {259–261}},\ \bibinfo {pages} {676 } (\bibinfo {year}
  {1999})}\BibitemShut {NoStop}%
\bibitem [{\citenamefont {Jaccard}\ \emph {et~al.}(1999)\citenamefont
  {Jaccard}, \citenamefont {Wilhelm}, \citenamefont {Alami-Yadri},\ and\
  \citenamefont {Vargoz}}]{Jaccard19991}%
  \BibitemOpen
  \bibfield  {author} {\bibinfo {author} {\bibfnamefont {D.}~\bibnamefont
  {Jaccard}}, \bibinfo {author} {\bibfnamefont {H.}~\bibnamefont {Wilhelm}},
  \bibinfo {author} {\bibfnamefont {K.}~\bibnamefont {Alami-Yadri}}, \ and\
  \bibinfo {author} {\bibfnamefont {E.}~\bibnamefont {Vargoz}},\ }\href
  {\doibase http://dx.doi.org/10.1016/S0921-4526(98)01069-2} {\bibfield
  {journal} {\bibinfo  {journal} {Physica B: Condensed Matter}\ }\textbf
  {\bibinfo {volume} {259–261}},\ \bibinfo {pages} {1 } (\bibinfo {year}
  {1999})}\BibitemShut {NoStop}%
\bibitem [{\citenamefont {Yuan}\ \emph {et~al.}(2003)\citenamefont {Yuan},
  \citenamefont {Grosche}, \citenamefont {Deppe}, \citenamefont {Geibel},
  \citenamefont {Sparn},\ and\ \citenamefont
  {Steglich}}]{Science.302.5653.2104}%
  \BibitemOpen
  \bibfield  {author} {\bibinfo {author} {\bibfnamefont {H.}~\bibnamefont
  {Yuan}}, \bibinfo {author} {\bibfnamefont {F.}~\bibnamefont {Grosche}},
  \bibinfo {author} {\bibfnamefont {M.}~\bibnamefont {Deppe}}, \bibinfo
  {author} {\bibfnamefont {C.}~\bibnamefont {Geibel}}, \bibinfo {author}
  {\bibfnamefont {G.}~\bibnamefont {Sparn}}, \ and\ \bibinfo {author}
  {\bibfnamefont {F.}~\bibnamefont {Steglich}},\ }\href@noop {} {\bibfield
  {journal} {\bibinfo  {journal} {Science}\ }\textbf {\bibinfo {volume}
  {302}},\ \bibinfo {pages} {2104} (\bibinfo {year} {2003})}\BibitemShut
  {NoStop}%
\bibitem [{\citenamefont {Holmes}\ \emph {et~al.}(2004)\citenamefont {Holmes},
  \citenamefont {Jaccard},\ and\ \citenamefont {Miyake}}]{PhysRevB.69.024508}%
  \BibitemOpen
  \bibfield  {author} {\bibinfo {author} {\bibfnamefont {A.~T.}\ \bibnamefont
  {Holmes}}, \bibinfo {author} {\bibfnamefont {D.}~\bibnamefont {Jaccard}}, \
  and\ \bibinfo {author} {\bibfnamefont {K.}~\bibnamefont {Miyake}},\ }\href
  {\doibase 10.1103/PhysRevB.69.024508} {\bibfield  {journal} {\bibinfo
  {journal} {Phys. Rev. B}\ }\textbf {\bibinfo {volume} {69}},\ \bibinfo
  {pages} {024508} (\bibinfo {year} {2004})}\BibitemShut {NoStop}%
\bibitem [{\citenamefont {Onishi}\ and\ \citenamefont {Miyake}(2000)}]{Onishi}%
  \BibitemOpen
  \bibfield  {author} {\bibinfo {author} {\bibfnamefont {Y.}~\bibnamefont
  {Onishi}}\ and\ \bibinfo {author} {\bibfnamefont {K.}~\bibnamefont
  {Miyake}},\ }\href {\doibase 10.1143/JPSJ.69.3955} {\bibfield  {journal}
  {\bibinfo  {journal} {J. Phys. Soc. Jpn.}\ }\textbf {\bibinfo {volume}
  {69}},\ \bibinfo {pages} {3955} (\bibinfo {year} {2000})}\BibitemShut
  {NoStop}%
\bibitem [{\citenamefont {Watanabe}\ \emph {et~al.}(2006)\citenamefont
  {Watanabe}, \citenamefont {Imada},\ and\ \citenamefont
  {Miyake}}]{JPSJ.75.043710}%
  \BibitemOpen
  \bibfield  {author} {\bibinfo {author} {\bibfnamefont {S.}~\bibnamefont
  {Watanabe}}, \bibinfo {author} {\bibfnamefont {M.}~\bibnamefont {Imada}}, \
  and\ \bibinfo {author} {\bibfnamefont {K.}~\bibnamefont {Miyake}},\ }\href
  {\doibase 10.1143/JPSJ.75.043710} {\bibfield  {journal} {\bibinfo  {journal}
  {J. Phys. Soc. Jpn.}\ }\textbf {\bibinfo {volume} {75}},\ \bibinfo {pages}
  {043710} (\bibinfo {year} {2006})}\BibitemShut {NoStop}%
\bibitem [{\citenamefont {Sugibayashi}\ \emph {et~al.}(2008)\citenamefont
  {Sugibayashi}, \citenamefont {Saiga},\ and\ \citenamefont
  {Hirashima}}]{JPSJ.77.024716}%
  \BibitemOpen
  \bibfield  {author} {\bibinfo {author} {\bibfnamefont {T.}~\bibnamefont
  {Sugibayashi}}, \bibinfo {author} {\bibfnamefont {Y.}~\bibnamefont {Saiga}},
  \ and\ \bibinfo {author} {\bibfnamefont {D.~S.}\ \bibnamefont {Hirashima}},\
  }\href {\doibase 10.1143/JPSJ.77.024716} {\bibfield  {journal} {\bibinfo
  {journal} {J. Phys. Soc. Jpn.}\ }\textbf {\bibinfo {volume} {77}},\ \bibinfo
  {pages} {024716} (\bibinfo {year} {2008})}\BibitemShut {NoStop}%
\bibitem [{\citenamefont {Kubo}(2011)}]{JPSJ.80.114711}%
  \BibitemOpen
  \bibfield  {author} {\bibinfo {author} {\bibfnamefont {K.}~\bibnamefont
  {Kubo}},\ }\href {\doibase 10.1143/JPSJ.80.114711} {\bibfield  {journal}
  {\bibinfo  {journal} {J. Phys. Soc. Jpn.}\ }\textbf {\bibinfo {volume}
  {80}},\ \bibinfo {pages} {114711} (\bibinfo {year} {2011})}\BibitemShut
  {NoStop}%
\bibitem [{\citenamefont {Kojima}\ and\ \citenamefont {Koga}(2013)}]{Kojima}%
  \BibitemOpen
  \bibfield  {author} {\bibinfo {author} {\bibfnamefont {Y.}~\bibnamefont
  {Kojima}}\ and\ \bibinfo {author} {\bibfnamefont {A.}~\bibnamefont {Koga}},\
  }\href@noop {} {\bibfield  {journal} {\bibinfo  {journal} {JPS Conf. Proc.}\
  }\textbf {\bibinfo {volume} {1}},\ \bibinfo {pages} {012106} (\bibinfo {year}
  {2013})}\BibitemShut {NoStop}%
\bibitem [{\citenamefont {Hagym\'asi}\ \emph {et~al.}(2013)\citenamefont
  {Hagym\'asi}, \citenamefont {Itai},\ and\ \citenamefont
  {S\'olyom}}]{PhysRevB.87.125146}%
  \BibitemOpen
  \bibfield  {author} {\bibinfo {author} {\bibfnamefont {I.}~\bibnamefont
  {Hagym\'asi}}, \bibinfo {author} {\bibfnamefont {K.}~\bibnamefont {Itai}}, \
  and\ \bibinfo {author} {\bibfnamefont {J.}~\bibnamefont {S\'olyom}},\ }\href
  {\doibase 10.1103/PhysRevB.87.125146} {\bibfield  {journal} {\bibinfo
  {journal} {Phys. Rev. B}\ }\textbf {\bibinfo {volume} {87}},\ \bibinfo
  {pages} {125146} (\bibinfo {year} {2013})}\BibitemShut {NoStop}%
\bibitem [{\citenamefont {Shinzaki}\ \emph {et~al.}(2016)\citenamefont
  {Shinzaki}, \citenamefont {Nasu},\ and\ \citenamefont
  {Koga}}]{JPCS2016012041}%
  \BibitemOpen
  \bibfield  {author} {\bibinfo {author} {\bibfnamefont {R.}~\bibnamefont
  {Shinzaki}}, \bibinfo {author} {\bibfnamefont {J.}~\bibnamefont {Nasu}}, \
  and\ \bibinfo {author} {\bibfnamefont {A.}~\bibnamefont {Koga}},\ }\href
  {http://stacks.iop.org/1742-6596/683/i=1/a=012041} {\bibfield  {journal}
  {\bibinfo  {journal} {J. Phys.: Conf. Ser.}\ }\textbf {\bibinfo {volume}
  {683}},\ \bibinfo {pages} {012041} (\bibinfo {year} {2016})}\BibitemShut
  {NoStop}%
\bibitem [{\citenamefont {Deguchi}\ \emph {et~al.}(2012)\citenamefont
  {Deguchi}, \citenamefont {Matsukawa}, \citenamefont {Sato}, \citenamefont
  {Hattori}, \citenamefont {Ishida}, \citenamefont {Takakura},\ and\
  \citenamefont {Ishimasa}}]{Nmat.12.1013}%
  \BibitemOpen
  \bibfield  {author} {\bibinfo {author} {\bibfnamefont {K.}~\bibnamefont
  {Deguchi}}, \bibinfo {author} {\bibfnamefont {S.}~\bibnamefont {Matsukawa}},
  \bibinfo {author} {\bibfnamefont {N.~K.}\ \bibnamefont {Sato}}, \bibinfo
  {author} {\bibfnamefont {T.}~\bibnamefont {Hattori}}, \bibinfo {author}
  {\bibfnamefont {K.}~\bibnamefont {Ishida}}, \bibinfo {author} {\bibfnamefont
  {H.}~\bibnamefont {Takakura}}, \ and\ \bibinfo {author} {\bibfnamefont
  {T.}~\bibnamefont {Ishimasa}},\ }\href@noop {} {\bibfield  {journal}
  {\bibinfo  {journal} {Nat. Mat.}\ }\textbf {\bibinfo {volume} {11}},\
  \bibinfo {pages} {1013} (\bibinfo {year} {2012})}\BibitemShut {NoStop}%
\bibitem [{\citenamefont {Matsukawa}\ \emph {et~al.}(2014)\citenamefont
  {Matsukawa}, \citenamefont {Tanaka}, \citenamefont {Nakayama}, \citenamefont
  {Deguchi}, \citenamefont {Imura}, \citenamefont {Takakura}, \citenamefont
  {Kashimoto}, \citenamefont {Ishimasa},\ and\ \citenamefont
  {Sato}}]{JPSJ.83.034705}%
  \BibitemOpen
  \bibfield  {author} {\bibinfo {author} {\bibfnamefont {S.}~\bibnamefont
  {Matsukawa}}, \bibinfo {author} {\bibfnamefont {K.}~\bibnamefont {Tanaka}},
  \bibinfo {author} {\bibfnamefont {M.}~\bibnamefont {Nakayama}}, \bibinfo
  {author} {\bibfnamefont {K.}~\bibnamefont {Deguchi}}, \bibinfo {author}
  {\bibfnamefont {K.}~\bibnamefont {Imura}}, \bibinfo {author} {\bibfnamefont
  {H.}~\bibnamefont {Takakura}}, \bibinfo {author} {\bibfnamefont
  {S.}~\bibnamefont {Kashimoto}}, \bibinfo {author} {\bibfnamefont
  {T.}~\bibnamefont {Ishimasa}}, \ and\ \bibinfo {author} {\bibfnamefont
  {N.~K.}\ \bibnamefont {Sato}},\ }\href {\doibase 10.7566/JPSJ.83.034705}
  {\bibfield  {journal} {\bibinfo  {journal} {J. Phys. Soc. Jpn.}\ }\textbf
  {\bibinfo {volume} {83}},\ \bibinfo {pages} {034705} (\bibinfo {year}
  {2014})}\BibitemShut {NoStop}%
\bibitem [{\citenamefont {Watanabe}\ and\ \citenamefont
  {Miyake}(2013)}]{JPSJ.82.083704}%
  \BibitemOpen
  \bibfield  {author} {\bibinfo {author} {\bibfnamefont {S.}~\bibnamefont
  {Watanabe}}\ and\ \bibinfo {author} {\bibfnamefont {K.}~\bibnamefont
  {Miyake}},\ }\href {\doibase 10.7566/JPSJ.82.083704} {\bibfield  {journal}
  {\bibinfo  {journal} {J. Phys. Soc. Jpn.}\ }\textbf {\bibinfo {volume}
  {82}},\ \bibinfo {pages} {083704} (\bibinfo {year} {2013})}\BibitemShut
  {NoStop}%
\bibitem [{\citenamefont {Andrade}\ \emph {et~al.}(2015)\citenamefont
  {Andrade}, \citenamefont {Jagannathan}, \citenamefont {Miranda},
  \citenamefont {Vojta},\ and\ \citenamefont
  {Dobrosavljevi\ifmmode~\acute{c}\else \'{c}\fi{}}}]{PhysRevLett.115.036403}%
  \BibitemOpen
  \bibfield  {author} {\bibinfo {author} {\bibfnamefont {E.~C.}\ \bibnamefont
  {Andrade}}, \bibinfo {author} {\bibfnamefont {A.}~\bibnamefont
  {Jagannathan}}, \bibinfo {author} {\bibfnamefont {E.}~\bibnamefont
  {Miranda}}, \bibinfo {author} {\bibfnamefont {M.}~\bibnamefont {Vojta}}, \
  and\ \bibinfo {author} {\bibfnamefont {V.}~\bibnamefont
  {Dobrosavljevi\ifmmode~\acute{c}\else \'{c}\fi{}}},\ }\href {\doibase
  10.1103/PhysRevLett.115.036403} {\bibfield  {journal} {\bibinfo  {journal}
  {Phys. Rev. Lett.}\ }\textbf {\bibinfo {volume} {115}},\ \bibinfo {pages}
  {036403} (\bibinfo {year} {2015})}\BibitemShut {NoStop}%
\bibitem [{\citenamefont {Takemura}\ \emph {et~al.}(2015)\citenamefont
  {Takemura}, \citenamefont {Takemori},\ and\ \citenamefont {Koga}}]{Takemura}%
  \BibitemOpen
  \bibfield  {author} {\bibinfo {author} {\bibfnamefont {S.}~\bibnamefont
  {Takemura}}, \bibinfo {author} {\bibfnamefont {N.}~\bibnamefont {Takemori}},
  \ and\ \bibinfo {author} {\bibfnamefont {A.}~\bibnamefont {Koga}},\ }\href
  {\doibase 10.1103/PhysRevB.91.165114} {\bibfield  {journal} {\bibinfo
  {journal} {Phys. Rev. B}\ }\textbf {\bibinfo {volume} {91}},\ \bibinfo
  {pages} {165114} (\bibinfo {year} {2015})}\BibitemShut {NoStop}%
\bibitem [{\citenamefont {Takemori}\ and\ \citenamefont
  {Koga}(2015)}]{Takemori}%
  \BibitemOpen
  \bibfield  {author} {\bibinfo {author} {\bibfnamefont {N.}~\bibnamefont
  {Takemori}}\ and\ \bibinfo {author} {\bibfnamefont {A.}~\bibnamefont
  {Koga}},\ }\href {\doibase 10.7566/JPSJ.84.023701} {\bibfield  {journal}
  {\bibinfo  {journal} {J. Phys. Soc. Jpn.}\ }\textbf {\bibinfo {volume}
  {84}},\ \bibinfo {pages} {023701} (\bibinfo {year} {2015})}\BibitemShut
  {NoStop}%
\bibitem [{\citenamefont {Kuramoto}(1983)}]{NCA1}%
  \BibitemOpen
  \bibfield  {author} {\bibinfo {author} {\bibfnamefont {Y.}~\bibnamefont
  {Kuramoto}},\ }\href {\doibase 10.1007/BF01578246} {\bibfield  {journal}
  {\bibinfo  {journal} {Z. Phys. B}\ }\textbf {\bibinfo {volume} {53}},\
  \bibinfo {pages} {37} (\bibinfo {year} {1983})}\BibitemShut {NoStop}%
\bibitem [{\citenamefont {Eckstein}\ and\ \citenamefont {Werner}(2010)}]{NCA2}%
  \BibitemOpen
  \bibfield  {author} {\bibinfo {author} {\bibfnamefont {M.}~\bibnamefont
  {Eckstein}}\ and\ \bibinfo {author} {\bibfnamefont {P.}~\bibnamefont
  {Werner}},\ }\href {\doibase 10.1103/PhysRevB.82.115115} {\bibfield
  {journal} {\bibinfo  {journal} {Phys. Rev. B}\ }\textbf {\bibinfo {volume}
  {82}},\ \bibinfo {pages} {115115} (\bibinfo {year} {2010})}\BibitemShut
  {NoStop}%
\bibitem [{\citenamefont {Georges}\ \emph {et~al.}(1996)\citenamefont
  {Georges}, \citenamefont {Kotliar}, \citenamefont {Krauth},\ and\
  \citenamefont {Rozenberg}}]{RevModPhys.68.13}%
  \BibitemOpen
  \bibfield  {author} {\bibinfo {author} {\bibfnamefont {A.}~\bibnamefont
  {Georges}}, \bibinfo {author} {\bibfnamefont {G.}~\bibnamefont {Kotliar}},
  \bibinfo {author} {\bibfnamefont {W.}~\bibnamefont {Krauth}}, \ and\ \bibinfo
  {author} {\bibfnamefont {M.~J.}\ \bibnamefont {Rozenberg}},\ }\href {\doibase
  10.1103/RevModPhys.68.13} {\bibfield  {journal} {\bibinfo  {journal} {Rev.
  Mod. Phys.}\ }\textbf {\bibinfo {volume} {68}},\ \bibinfo {pages} {13}
  (\bibinfo {year} {1996})}\BibitemShut {NoStop}%
\bibitem [{\citenamefont {Pruschke}\ \emph {et~al.}(1995)\citenamefont
  {Pruschke}, \citenamefont {Jarrell},\ and\ \citenamefont
  {Freericks}}]{AdvPhys.44.187}%
  \BibitemOpen
  \bibfield  {author} {\bibinfo {author} {\bibfnamefont {T.}~\bibnamefont
  {Pruschke}}, \bibinfo {author} {\bibfnamefont {M.}~\bibnamefont {Jarrell}}, \
  and\ \bibinfo {author} {\bibfnamefont {J.}~\bibnamefont {Freericks}},\
  }\href@noop {} {\bibfield  {journal} {\bibinfo  {journal} {Adv. Phys.}\
  }\textbf {\bibinfo {volume} {44}},\ \bibinfo {pages} {187} (\bibinfo {year}
  {1995})}\BibitemShut {NoStop}%
\bibitem [{\citenamefont {M{\"u}ller-Hartmann}()}]{ZPhysB.74.507}%
  \BibitemOpen
  \bibfield  {author} {\bibinfo {author} {\bibfnamefont {E.}~\bibnamefont
  {M{\"u}ller-Hartmann}},\ }\href {\doibase 10.1007/BF01311397} {\bibfield
  {journal} {\bibinfo  {journal} {Z. Phys. B}\ }\textbf {\bibinfo {volume}
  {74}},\ \bibinfo {pages} {507}}\BibitemShut {NoStop}%
\bibitem [{\citenamefont {Metzner}\ and\ \citenamefont
  {Vollhardt}(1989)}]{PhysRevLett.62.324}%
  \BibitemOpen
  \bibfield  {author} {\bibinfo {author} {\bibfnamefont {W.}~\bibnamefont
  {Metzner}}\ and\ \bibinfo {author} {\bibfnamefont {D.}~\bibnamefont
  {Vollhardt}},\ }\href {\doibase 10.1103/PhysRevLett.62.324} {\bibfield
  {journal} {\bibinfo  {journal} {Phys. Rev. Lett.}\ }\textbf {\bibinfo
  {volume} {62}},\ \bibinfo {pages} {324} (\bibinfo {year} {1989})}\BibitemShut
  {NoStop}%
\bibitem [{\citenamefont {Grewe}\ and\ \citenamefont {Steglich}()}]{Grewe}%
  \BibitemOpen
  \bibfield  {author} {\bibinfo {author} {\bibfnamefont {N.}~\bibnamefont
  {Grewe}}\ and\ \bibinfo {author} {\bibfnamefont {F.}~\bibnamefont
  {Steglich}},\ }\href@noop {} {\bibfield  {journal} {\bibinfo  {journal} {in
  {\it{Handbook on the Physics and Chemistry of Rare Earths}}, edited by K. A.
  Gschneidner, Jr. and L. Eyring (North-Holland, Amsterdam, 1991)}\ }\textbf
  {\bibinfo {volume} {14}},\ \bibinfo {pages} {343}}\BibitemShut {NoStop}%
\bibitem [{\citenamefont {Ohashi}\ \emph {et~al.}(2004)\citenamefont {Ohashi},
  \citenamefont {Koga}, \citenamefont {Suga},\ and\ \citenamefont
  {Kawakami}}]{ohashi_field-induced}%
  \BibitemOpen
  \bibfield  {author} {\bibinfo {author} {\bibfnamefont {T.}~\bibnamefont
  {Ohashi}}, \bibinfo {author} {\bibfnamefont {A.}~\bibnamefont {Koga}},
  \bibinfo {author} {\bibfnamefont {S.}~\bibnamefont {Suga}}, \ and\ \bibinfo
  {author} {\bibfnamefont {N.}~\bibnamefont {Kawakami}},\ }\href {\doibase
  10.1103/PhysRevB.70.245104} {\bibfield  {journal} {\bibinfo  {journal} {Phys.
  Rev. B}\ }\textbf {\bibinfo {volume} {70}},\ \bibinfo {pages} {245104}
  (\bibinfo {year} {2004})}\BibitemShut {NoStop}%
\bibitem [{\citenamefont {Sordi}\ \emph {et~al.}(2007)\citenamefont {Sordi},
  \citenamefont {Amaricci},\ and\ \citenamefont
  {Rozenberg}}]{PhysRevLett.99.196403}%
  \BibitemOpen
  \bibfield  {author} {\bibinfo {author} {\bibfnamefont {G.}~\bibnamefont
  {Sordi}}, \bibinfo {author} {\bibfnamefont {A.}~\bibnamefont {Amaricci}}, \
  and\ \bibinfo {author} {\bibfnamefont {M.~J.}\ \bibnamefont {Rozenberg}},\
  }\href {\doibase 10.1103/PhysRevLett.99.196403} {\bibfield  {journal}
  {\bibinfo  {journal} {Phys. Rev. Lett.}\ }\textbf {\bibinfo {volume} {99}},\
  \bibinfo {pages} {196403} (\bibinfo {year} {2007})}\BibitemShut {NoStop}%
\bibitem [{\citenamefont {Koga}\ and\ \citenamefont
  {Werner}(2010)}]{koga_2010}%
  \BibitemOpen
  \bibfield  {author} {\bibinfo {author} {\bibfnamefont {A.}~\bibnamefont
  {Koga}}\ and\ \bibinfo {author} {\bibfnamefont {P.}~\bibnamefont {Werner}},\
  }\href {\doibase 10.1143/JPSJ.79.114401} {\bibfield  {journal} {\bibinfo
  {journal} {J. Phys. Soc. Jpn.}\ }\textbf {\bibinfo {volume} {79}},\ \bibinfo
  {pages} {114401} (\bibinfo {year} {2010})}\BibitemShut {NoStop}%
\bibitem [{\citenamefont {Levine}\ and\ \citenamefont
  {Steinhardt}(1984)}]{PhysRevLett.53.2477}%
  \BibitemOpen
  \bibfield  {author} {\bibinfo {author} {\bibfnamefont {D.}~\bibnamefont
  {Levine}}\ and\ \bibinfo {author} {\bibfnamefont {P.~J.}\ \bibnamefont
  {Steinhardt}},\ }\href {\doibase 10.1103/PhysRevLett.53.2477} {\bibfield
  {journal} {\bibinfo  {journal} {Phys. Rev. Lett.}\ }\textbf {\bibinfo
  {volume} {53}},\ \bibinfo {pages} {2477} (\bibinfo {year}
  {1984})}\BibitemShut {NoStop}%
\bibitem [{\citenamefont {Potthoff}\ and\ \citenamefont
  {Nolting}(1999)}]{Potthoff}%
  \BibitemOpen
  \bibfield  {author} {\bibinfo {author} {\bibfnamefont {M.}~\bibnamefont
  {Potthoff}}\ and\ \bibinfo {author} {\bibfnamefont {W.}~\bibnamefont
  {Nolting}},\ }\href {\doibase 10.1103/PhysRevB.59.2549} {\bibfield  {journal}
  {\bibinfo  {journal} {Phys. Rev. B}\ }\textbf {\bibinfo {volume} {59}},\
  \bibinfo {pages} {2549} (\bibinfo {year} {1999})}\BibitemShut {NoStop}%
\bibitem [{\citenamefont {Okamoto}\ and\ \citenamefont
  {Millis}(2004)}]{Okamoto}%
  \BibitemOpen
  \bibfield  {author} {\bibinfo {author} {\bibfnamefont {S.}~\bibnamefont
  {Okamoto}}\ and\ \bibinfo {author} {\bibfnamefont {A.~J.}\ \bibnamefont
  {Millis}},\ }\href {\doibase 10.1103/PhysRevB.70.241104} {\bibfield
  {journal} {\bibinfo  {journal} {Phys. Rev. B}\ }\textbf {\bibinfo {volume}
  {70}},\ \bibinfo {pages} {241104} (\bibinfo {year} {2004})}\BibitemShut
  {NoStop}%
\bibitem [{\citenamefont {Peters}\ \emph {et~al.}(2013)\citenamefont {Peters},
  \citenamefont {Tada},\ and\ \citenamefont {Kawakami}}]{Peters}%
  \BibitemOpen
  \bibfield  {author} {\bibinfo {author} {\bibfnamefont {R.}~\bibnamefont
  {Peters}}, \bibinfo {author} {\bibfnamefont {Y.}~\bibnamefont {Tada}}, \ and\
  \bibinfo {author} {\bibfnamefont {N.}~\bibnamefont {Kawakami}},\ }\href
  {\doibase 10.1103/PhysRevB.88.155134} {\bibfield  {journal} {\bibinfo
  {journal} {Phys. Rev. B}\ }\textbf {\bibinfo {volume} {88}},\ \bibinfo
  {pages} {155134} (\bibinfo {year} {2013})}\BibitemShut {NoStop}%
\bibitem [{\citenamefont {Snoek}\ \emph {et~al.}(2008)\citenamefont {Snoek},
  \citenamefont {Titvinidze}, \citenamefont {T\"oke}, \citenamefont {Byczuk},\
  and\ \citenamefont {Hofstetter}}]{Snoek}%
  \BibitemOpen
  \bibfield  {author} {\bibinfo {author} {\bibfnamefont {M.}~\bibnamefont
  {Snoek}}, \bibinfo {author} {\bibfnamefont {I.}~\bibnamefont {Titvinidze}},
  \bibinfo {author} {\bibfnamefont {C.}~\bibnamefont {T\"oke}}, \bibinfo
  {author} {\bibfnamefont {K.}~\bibnamefont {Byczuk}}, \ and\ \bibinfo {author}
  {\bibfnamefont {W.}~\bibnamefont {Hofstetter}},\ }\href
  {http://stacks.iop.org/1367-2630/10/i=9/a=093008} {\bibfield  {journal}
  {\bibinfo  {journal} {New J. Phys.}\ }\textbf {\bibinfo {volume} {10}},\
  \bibinfo {pages} {093008} (\bibinfo {year} {2008})}\BibitemShut {NoStop}%
\bibitem [{\citenamefont {Helmes}\ \emph {et~al.}(2008)\citenamefont {Helmes},
  \citenamefont {Costi},\ and\ \citenamefont {Rosch}}]{Helmes}%
  \BibitemOpen
  \bibfield  {author} {\bibinfo {author} {\bibfnamefont {R.~W.}\ \bibnamefont
  {Helmes}}, \bibinfo {author} {\bibfnamefont {T.~A.}\ \bibnamefont {Costi}}, \
  and\ \bibinfo {author} {\bibfnamefont {A.}~\bibnamefont {Rosch}},\ }\href
  {\doibase 10.1103/PhysRevLett.100.056403} {\bibfield  {journal} {\bibinfo
  {journal} {Phys. Rev. Lett.}\ }\textbf {\bibinfo {volume} {100}},\ \bibinfo
  {pages} {056403} (\bibinfo {year} {2008})}\BibitemShut {NoStop}%
\bibitem [{\citenamefont {Koga}\ \emph {et~al.}(2008)\citenamefont {Koga},
  \citenamefont {Higashiyama}, \citenamefont {Inaba}, \citenamefont {Suga},\
  and\ \citenamefont {Kawakami}}]{KogaOL}%
  \BibitemOpen
  \bibfield  {author} {\bibinfo {author} {\bibfnamefont {A.}~\bibnamefont
  {Koga}}, \bibinfo {author} {\bibfnamefont {T.}~\bibnamefont {Higashiyama}},
  \bibinfo {author} {\bibfnamefont {K.}~\bibnamefont {Inaba}}, \bibinfo
  {author} {\bibfnamefont {S.}~\bibnamefont {Suga}}, \ and\ \bibinfo {author}
  {\bibfnamefont {N.}~\bibnamefont {Kawakami}},\ }\href {\doibase
  10.1143/JPSJ.77.073602} {\bibfield  {journal} {\bibinfo  {journal} {J. Phys.
  Soc. Jpn.}\ }\textbf {\bibinfo {volume} {77}},\ \bibinfo {pages} {073602}
  (\bibinfo {year} {2008})}\BibitemShut {NoStop}%
\bibitem [{\citenamefont {Koga}\ \emph {et~al.}(2009)\citenamefont {Koga},
  \citenamefont {Higashiyama}, \citenamefont {Inaba}, \citenamefont {Suga},\
  and\ \citenamefont {Kawakami}}]{KogaOL2}%
  \BibitemOpen
  \bibfield  {author} {\bibinfo {author} {\bibfnamefont {A.}~\bibnamefont
  {Koga}}, \bibinfo {author} {\bibfnamefont {T.}~\bibnamefont {Higashiyama}},
  \bibinfo {author} {\bibfnamefont {K.}~\bibnamefont {Inaba}}, \bibinfo
  {author} {\bibfnamefont {S.}~\bibnamefont {Suga}}, \ and\ \bibinfo {author}
  {\bibfnamefont {N.}~\bibnamefont {Kawakami}},\ }\href {\doibase
  10.1103/PhysRevA.79.013607} {\bibfield  {journal} {\bibinfo  {journal} {Phys.
  Rev. A}\ }\textbf {\bibinfo {volume} {79}},\ \bibinfo {pages} {013607}
  (\bibinfo {year} {2009})}\BibitemShut {NoStop}%
\bibitem [{\citenamefont {Tada}\ \emph {et~al.}(2012)\citenamefont {Tada},
  \citenamefont {Peters}, \citenamefont {Oshikawa}, \citenamefont {Koga},
  \citenamefont {Kawakami},\ and\ \citenamefont {Fujimoto}}]{Tada}%
  \BibitemOpen
  \bibfield  {author} {\bibinfo {author} {\bibfnamefont {Y.}~\bibnamefont
  {Tada}}, \bibinfo {author} {\bibfnamefont {R.}~\bibnamefont {Peters}},
  \bibinfo {author} {\bibfnamefont {M.}~\bibnamefont {Oshikawa}}, \bibinfo
  {author} {\bibfnamefont {A.}~\bibnamefont {Koga}}, \bibinfo {author}
  {\bibfnamefont {N.}~\bibnamefont {Kawakami}}, \ and\ \bibinfo {author}
  {\bibfnamefont {S.}~\bibnamefont {Fujimoto}},\ }\href {\doibase
  10.1103/PhysRevB.85.165138} {\bibfield  {journal} {\bibinfo  {journal} {Phys.
  Rev. B}\ }\textbf {\bibinfo {volume} {85}},\ \bibinfo {pages} {165138}
  (\bibinfo {year} {2012})}\BibitemShut {NoStop}%
\bibitem [{\citenamefont {Werner}\ \emph {et~al.}(2006)\citenamefont {Werner},
  \citenamefont {Comanac}, \citenamefont {de' Medici}, \citenamefont {Troyer},\
  and\ \citenamefont {Millis}}]{PhysRevLett.97.076405}%
  \BibitemOpen
  \bibfield  {author} {\bibinfo {author} {\bibfnamefont {P.}~\bibnamefont
  {Werner}}, \bibinfo {author} {\bibfnamefont {A.}~\bibnamefont {Comanac}},
  \bibinfo {author} {\bibfnamefont {L.}~\bibnamefont {de' Medici}}, \bibinfo
  {author} {\bibfnamefont {M.}~\bibnamefont {Troyer}}, \ and\ \bibinfo {author}
  {\bibfnamefont {A.~J.}\ \bibnamefont {Millis}},\ }\href {\doibase
  10.1103/PhysRevLett.97.076405} {\bibfield  {journal} {\bibinfo  {journal}
  {Phys. Rev. Lett.}\ }\textbf {\bibinfo {volume} {97}},\ \bibinfo {pages}
  {076405} (\bibinfo {year} {2006})}\BibitemShut {NoStop}%
\bibitem [{\citenamefont {Werner}\ and\ \citenamefont
  {Millis}(2006)}]{PhysRevB.74.155107}%
  \BibitemOpen
  \bibfield  {author} {\bibinfo {author} {\bibfnamefont {P.}~\bibnamefont
  {Werner}}\ and\ \bibinfo {author} {\bibfnamefont {A.~J.}\ \bibnamefont
  {Millis}},\ }\href {\doibase 10.1103/PhysRevB.74.155107} {\bibfield
  {journal} {\bibinfo  {journal} {Phys. Rev. B}\ }\textbf {\bibinfo {volume}
  {74}},\ \bibinfo {pages} {155107} (\bibinfo {year} {2006})}\BibitemShut
  {NoStop}%
\bibitem [{\citenamefont {Gull}\ \emph {et~al.}(2011)\citenamefont {Gull},
  \citenamefont {Millis}, \citenamefont {Lichtenstein}, \citenamefont
  {Rubtsov}, \citenamefont {Troyer},\ and\ \citenamefont
  {Werner}}]{RevModPhys.83.349}%
  \BibitemOpen
  \bibfield  {author} {\bibinfo {author} {\bibfnamefont {E.}~\bibnamefont
  {Gull}}, \bibinfo {author} {\bibfnamefont {A.~J.}\ \bibnamefont {Millis}},
  \bibinfo {author} {\bibfnamefont {A.~I.}\ \bibnamefont {Lichtenstein}},
  \bibinfo {author} {\bibfnamefont {A.~N.}\ \bibnamefont {Rubtsov}}, \bibinfo
  {author} {\bibfnamefont {M.}~\bibnamefont {Troyer}}, \ and\ \bibinfo {author}
  {\bibfnamefont {P.}~\bibnamefont {Werner}},\ }\href {\doibase
  10.1103/RevModPhys.83.349} {\bibfield  {journal} {\bibinfo  {journal} {Rev.
  Mod. Phys.}\ }\textbf {\bibinfo {volume} {83}},\ \bibinfo {pages} {349}
  (\bibinfo {year} {2011})}\BibitemShut {NoStop}%
\bibitem [{\citenamefont {Zhang}\ \emph {et~al.}(1993)\citenamefont {Zhang},
  \citenamefont {Rozenberg},\ and\ \citenamefont {Kotliar}}]{IPT}%
  \BibitemOpen
  \bibfield  {author} {\bibinfo {author} {\bibfnamefont {X.~Y.}\ \bibnamefont
  {Zhang}}, \bibinfo {author} {\bibfnamefont {M.~J.}\ \bibnamefont
  {Rozenberg}}, \ and\ \bibinfo {author} {\bibfnamefont {G.}~\bibnamefont
  {Kotliar}},\ }\href {\doibase 10.1103/PhysRevLett.70.1666} {\bibfield
  {journal} {\bibinfo  {journal} {Phys. Rev. Lett.}\ }\textbf {\bibinfo
  {volume} {70}},\ \bibinfo {pages} {1666} (\bibinfo {year}
  {1993})}\BibitemShut {NoStop}%
\bibitem [{\citenamefont {Bauer}\ \emph {et~al.}(2011)\citenamefont {Bauer},
  \citenamefont {Carr}, \citenamefont {Evertz}, \citenamefont {Feiguin},
  \citenamefont {Freire}, \citenamefont {Fuchs}, \citenamefont {Gamper},
  \citenamefont {Gukelberger}, \citenamefont {Gull}, \citenamefont {Guertler},
  \citenamefont {{A Hehn}}, \citenamefont {Igarashi}, \citenamefont {Isakov},
  \citenamefont {Koop}, \citenamefont {Ma}, \citenamefont {Mates},
  \citenamefont {Matsuo}, \citenamefont {Parcollet}, \citenamefont
  {Paw{\l}owski}, \citenamefont {Picon}, \citenamefont {{L Pollet}},
  \citenamefont {Santos}, \citenamefont {Scarola}, \citenamefont
  {Schollw\"{o}ck}, \citenamefont {Silva}, \citenamefont {Surer}, \citenamefont
  {Todo}, \citenamefont {Trebst}, \citenamefont {Troyer}, \citenamefont {Wall},
  \citenamefont {{P Werner}},\ and\ \citenamefont {Wessel}}]{alps2}%
  \BibitemOpen
  \bibfield  {author} {\bibinfo {author} {\bibfnamefont {B.}~\bibnamefont
  {Bauer}}, \bibinfo {author} {\bibfnamefont {L.~D.}\ \bibnamefont {Carr}},
  \bibinfo {author} {\bibfnamefont {H.~G.}\ \bibnamefont {Evertz}}, \bibinfo
  {author} {\bibfnamefont {A.}~\bibnamefont {Feiguin}}, \bibinfo {author}
  {\bibfnamefont {J.}~\bibnamefont {Freire}}, \bibinfo {author} {\bibfnamefont
  {S.}~\bibnamefont {Fuchs}}, \bibinfo {author} {\bibfnamefont
  {L.}~\bibnamefont {Gamper}}, \bibinfo {author} {\bibfnamefont
  {J.}~\bibnamefont {Gukelberger}}, \bibinfo {author} {\bibfnamefont
  {E.}~\bibnamefont {Gull}}, \bibinfo {author} {\bibfnamefont {S.}~\bibnamefont
  {Guertler}}, \bibinfo {author} {\bibnamefont {{A Hehn}}}, \bibinfo {author}
  {\bibfnamefont {R.}~\bibnamefont {Igarashi}}, \bibinfo {author}
  {\bibfnamefont {S.~V.}\ \bibnamefont {Isakov}}, \bibinfo {author}
  {\bibfnamefont {D.}~\bibnamefont {Koop}}, \bibinfo {author} {\bibfnamefont
  {P.~N.}\ \bibnamefont {Ma}}, \bibinfo {author} {\bibfnamefont
  {P.}~\bibnamefont {Mates}}, \bibinfo {author} {\bibfnamefont
  {H.}~\bibnamefont {Matsuo}}, \bibinfo {author} {\bibfnamefont
  {O.}~\bibnamefont {Parcollet}}, \bibinfo {author} {\bibfnamefont
  {G.}~\bibnamefont {Paw{\l}owski}}, \bibinfo {author} {\bibfnamefont {J.~D.}\
  \bibnamefont {Picon}}, \bibinfo {author} {\bibnamefont {{L Pollet}}},
  \bibinfo {author} {\bibfnamefont {E.}~\bibnamefont {Santos}}, \bibinfo
  {author} {\bibfnamefont {V.~W.}\ \bibnamefont {Scarola}}, \bibinfo {author}
  {\bibfnamefont {U.}~\bibnamefont {Schollw\"{o}ck}}, \bibinfo {author}
  {\bibfnamefont {C.}~\bibnamefont {Silva}}, \bibinfo {author} {\bibfnamefont
  {B.}~\bibnamefont {Surer}}, \bibinfo {author} {\bibfnamefont
  {S.}~\bibnamefont {Todo}}, \bibinfo {author} {\bibfnamefont {S.}~\bibnamefont
  {Trebst}}, \bibinfo {author} {\bibfnamefont {M.}~\bibnamefont {Troyer}},
  \bibinfo {author} {\bibfnamefont {M.~L.}\ \bibnamefont {Wall}}, \bibinfo
  {author} {\bibnamefont {{P Werner}}}, \ and\ \bibinfo {author} {\bibfnamefont
  {S.}~\bibnamefont {Wessel}},\ }\href {\doibase
  10.1088/1742-5468/2011/05/P05001} {\bibfield  {journal} {\bibinfo  {journal}
  {J. Stat. Mech. Theory Exp.}\ }\textbf {\bibinfo {volume} {2011}},\ \bibinfo
  {pages} {P05001} (\bibinfo {year} {2011})}\BibitemShut {NoStop}%
\end{thebibliography}%

\end{document}